\documentclass[prd,onecolumn,preprintnumbers,nofootinbib]{revtex4}
\usepackage{graphicx}

\usepackage[plainpages=false, colorlinks=true, anchorcolor=blue, linkcolor=blue, citecolor=blue, bookmarks=false]{hyperref}
\usepackage{color}
\newcommand{\rthis}[1]{\textcolor{black}{#1}}
\usepackage{float}
\usepackage{amsfonts,amsmath,amssymb}
\usepackage{natbib}
\usepackage{array}
\usepackage{graphicx}
\usepackage{pdfpages}
\usepackage{booktabs}
\usepackage{hhline}
\usepackage[font=footnotesize]{caption}
\usepackage{multirow}
\usepackage{url}
\begin{document}
\newcolumntype{P}[1]{>{\centering\arraybackslash}p{#1}}
\pdfoutput=1
\newcommand{\jcap}{JCAP}
\newcommand{\araa}{Annual Review of Astron. and Astrophys.}
\newcommand{\apss}{Astrophysics and Space Sciences}
\newcommand{\aj}{Astron. J. }
\newcommand{\mnras}{MNRAS}
\newcommand{\physrep}{Physics Reports}
\newcommand{\apjl}{Astrophys. J. Lett.}
\newcommand{\apjs}{Astrophys. J. Suppl. Ser.}
\newcommand{\aap}{Astron. \& Astrophys.}
\newcommand{\pasa}{PASA}
\renewcommand{\arraystretch}{2.5}
\title{Bayesian evidence for spectral lag transition due to Lorentz Invariance Violation for 32 Fermi/GBM Gamma-ray Bursts}
\author{Vibhavasu \surname{Pasumarti}}
\altaffiliation{E-mail:ep20btech11015@iith.ac.in}

\author{Shantanu \surname{Desai}}
\altaffiliation{E-mail: shntn05@gmail.com}

\begin{abstract}
We use the spectral lag data of 32 long GRBs detected by Fermi/GBM, which has been recently collated in ~\cite{Liu_2022} to  quantify  the statistical significance of a transition in the spectral lag data based on Lorentz invariance violation (LIV)  (for both sub-luminal and super-luminal propagation) using Bayesian model selection.
 We use two different parametric functions to model the null hypothesis of only intrinsic emission: a smooth broken power law model (SBPL) (proposed in ~\cite{Liu_2022}) as well as a simple power law model, which has been widely used before in literature.  We find that for sub-luminal propagation,  when we use  the SBPL model as the null hypothesis, \rthis{five} GRBs show ``decisive evidence'' based on Jeffreys' scale for linear LIV and  quadratic LIV. When we use  the simple power-law model as the null hypothesis, we find that \rthis{10 and 9} GRBs show  Bayesian ``decisive evidence'' for linear and quadratic LIV, respectively.  However these results should not be construed as evidence for LIV, as they would be in  conflict with the most stringent upper limits.  When we did a test for super-luminal LIV, we find that only four and two  GRBs show Bayesian ``decisive evidence'' for linear and quadratic LIV, respectively,  assuming a simple power law for the intrinsic emission. \rthis{When we use the  SBPL model, one GRB shows Bayesian ``decisive evidence'' for linear and quadratic LIV.}  This underscores the importance of adequately modelling the intrinsic emission while  obtaining constraints on  LIV using spectral lags, since inadequate modelling could masquerade as a signature of LIV. 
\end{abstract}

\affiliation{Dept  of Physics, IIT Hyderabad,  Kandi, Telangana-502284, India}

\maketitle
\section{Introduction}
In various  theoretical scenarios beyond the Standard Model of Particle Physics, Lorentz Invariance is not  an exact symmetry at energies close to the Planck scale ($E_{pl} \sim 10^{19}$ GeV), and
the speed of light $v(E)$ varies as a function of energy according to~\cite{GAA98}: 
\begin{equation}
    v(E) = c\left[1 - s_{\pm}\frac{n+1}{2} \left(\frac{E}{E_{QG}}\right)^n\right],
    \label{eq:vE}
\end{equation}
where $s_{\pm} = \pm 1$ corresponds to either sub-luminal ($s_{\pm}=+1$) or super-luminal ($s_{\pm}=-1$)  Lorentz Invariance Violation (LIV); $E_{QG}$ denotes the energy scale where LIV effects dominate, and $n$ represents the order of the modification of the photon group velocity. In all LIV searches in literature,  the series expansion is usually restricted  to linear ($n=1$) or quadratic corrections ($n=2$).  Both linear and quadratic LIV models are predicted by different theoretical approaches~\cite{Addazi21}. 

For more than two decades, Gamma-Ray Bursts (GRBs) have been a very powerful probe of LIV searches.  
GRBs are single-shot explosions located at cosmological distances, which were first detected in 1960s  and have been observed over ten  decades in energies from  keV to over 10~TeV range~\cite{Kumar,WuGRBreview}.  They are located at cosmological distances, although a distinct time-dilation signature in the light curves is yet to be demonstrated~\cite{Singh}. GRBs are traditionally divided into two categories based on their durations, with  long (short) GRBs lasting more (less) than two seconds~\cite{Kouveliotou}.
Long GRBs are usually associated with core-collapse SN~\cite{Woosley} and short GRBs with neutron star mergers~\cite{Nakar}. There are however many exceptions to this conventonal dichotomy, and many claims for additional GRB sub-classes have also been made~\cite{Kulkarni,Bhave} (and references therein).

The observable used in almost all the  LIV searches with GRBs consists of spectral lags,  defined as the arrival time difference  between high energy and low energy photons, and is positive if the high energy photons precede the low energy ones. Searches for LIV with spectral lags  have been done using  single lags from different GRBs (for example ~\cite{Ellis}), multiple spectral lags from the same GRB (GRB 1606025B, GRB 1901114C,  GRB 1905030A)~\cite{Wei,Du,Ganguly,Liang23}, as well as stacking spectral lags from multiple  GRBs~\cite{Agrawal_2021}. A comprehensive uptodate review of all searches for LIV using GRB spectral lags can be found in  ~\cite{WeiWu2021,WeiWu2,Desairev}. 

Most recently, Liu et al~\cite{Liu_2022} (L22, hereafter) did a comprehensive study of LIV (assuming sub-luminal propagation with $s_{\pm}=+1$) using spectral lags of 32 long GRBs detected by Fermi/GBM. Most of  the GRBs studied in L22 had a turn-over in the spectral lag data. The intrinsic model which they used consisted of a smooth broken power law as a function of energy. L22 then obtained limits on LIV for both a linear and quadratic model of LIV  for each of the 32 GRBs. The characteristic limits they obtained were  $E_{QG} \gtrsim 1.5 \times 10^{14}$ GeV and $E_{QG} \gtrsim 8 \times 10^{5}$ GeV for linear and quadratic LIV, respectively.

In this work, we supplement  the analysis in L22, by calculating the statistical  significance of spectral lag transitions assuming linear and quadratic models of LIV compared to only the intrinsic astrophysical emission using Bayesian model comparison, similar to  our past works~\cite{Ganguly,Agrawal_2021,Desai23}. We also test the efficacy of the more prosaic power law model which has been extensively used previously in mainly LIV searches starting from Ref.~\cite{Wei} and compare the two sets of results.

The outline of this manuscript is as follows. We discuss the GRB dataset used for this work  and the analysis procedure in Sect.~\ref{sec:dataset}.  In Sect.~\ref{sec:analysis}, we recap our analysis procedure. In Sect.~\ref{sec:bayes}, we provide a   brief primer on Bayesian model comparison.
Our results are outlined in Sect.~\ref{sec:results}.  We conclude in  Sect.~\ref{sec:conclusions}.

\section{Dataset}
\label{sec:dataset}
We briefly describe the data analysis procedure in L22, where more details can be found. The sample chosen in L22 consists of 32  long GRBs in the redshift range $z \in [0.54 ,  4.35]$ chosen from the Fermi/GBM catalogue~\cite{VonKienlin}
We use the spectral lag data of 32 GRBs collated in L22, which has been kindly provided to us. These data consist of the observed energy (in keV) \rthis{along with its  associated uncertainty},  and the corresponding spectral lags along with their associated uncertainty. The extraction of light curves  followed by the  spectral lag calculation  were done  using the methods described in ~\cite{Yang20,Zhang21,Wang21}.  These spectral lags were calculated using distinct energy bands as well as bin sizes.  The data for all the 32 GRBs including bin size, number of energy bands, energy range  used for spectral lag calculation, redshift,  and time interval can be found in Table 1 of L22. Each GRB contains about 15-20 spectral lag data in the 10-1000 keV energy range. We note that the spectral lags for  two  of the GRBs in this catalog, namely GRB 1606025B and   GRB 190114C have been reported before and used to search for  LIV~\cite{Wei,Du}. However, in the previous works the spectral lag data for GRB 1606025B and   GRB 190114C were binned in energy, whereas the L22 spectral lag data  have been provided   at specific energies.

\section{Analysis}
\label{sec:analysis}
In this work, we follow the same analysis procedure as L22 to predict the spectral lag of GRBs. The observed spectral lag is given by:
\begin{equation}
    \Delta t_{obs} = \Delta t_{int} + \Delta t_{LIV},
    \label{eq:totaldelta}
\end{equation}
where $\Delta t_{int}$ is the intrinsic lag from the GRB radiation which is purely astrophysical and $\Delta t_{LIV}$ is the contribution from LIV.

Since the exact physical process for the intrinsic lag is not yet found and could vary according based on the GRB,  we use two different models for the intrinsic emission. The first model which we use is the  Smoothly broken power-law (SBPL) to model, proposed in L22 and can be written as:
\begin{equation}
    \Delta t_{int} = \zeta \left(\dfrac{E - E_0}{E_b}\right)^{\alpha_1}\left(\dfrac{1}{2}\Biggl[1 + \left(\dfrac{E - E_0}{E_b}\right)^{\frac{1}{\mu}}\Biggr]\right)^{(\alpha_2 - \alpha_1)\mu} ,
    \label{eq:int}
\end{equation}
where $\zeta$ is the normalization parameter, $E_b$ is the transition energy, $\alpha_1$ and $\alpha_2$ are the slopes before and after $E_b$, and $\mu$ is the transition smoothness. This SBPL transforms into a single power law for $\alpha_1 = \alpha_2$. The SBPL enables us to account for the negative lags observed in the data~\cite{Liu_2022}.

We also use the power law model first proposed in ~\cite{Wei}, which has been used in a number of works on LIV~\cite{Wei,Wei2,Wei17,Ganguly,Pan,Du,Agrawal_2021,Desai23}, and was motivated based on the analysis of single-pulse properties of about 50 GRBs~\cite{Shao}: 
\begin{equation} 
\Delta t_{int}  = (1+z) \tau \Big[ \Big(\frac{E}{keV}\Big)^{\alpha}-\Big(\frac{E_0}{keV}\Big)^{\alpha}\Big],
\label{eq:intwei}
\end{equation}

The model for the spectral lag originating from LIV is same as the one used in ~\cite{Jacob} and is given by:
\begin{equation}
     \Delta t_{\mathrm{LIV}}=-\frac{1+n}{2 H_{0}} \frac{E^{n}-E_{0}^{n}}{E_{\mathrm{QG}, n}^{n}} \int_{0}^{z} \frac{\left(1+z^{\prime}\right)^{n} d z^{\prime}}{\sqrt{\Omega_{\mathrm{m}}\left(1+z^{\prime}\right)^{3}+\Omega_{\Lambda}}},
\label{eq:spectrallag}     
\end{equation}
Note the above equation assumes that the expansion history of the universe is described by the $\Lambda$CDM model. Other expansion histories as well as  non-parametric methods to model the expansion histories have also been considered~\cite{Biesiada,Du,Agrawal_2021,Desai23} (and references therein). However, it has been found that the final results do not change much compared to using $\Lambda$CDM expansion history~\cite{Desai23}. Therefore, for this work we use Eq.~\ref{eq:spectrallag} to calculate the lag due to LIV. 
The cosmological parameters which we use are $H_0 = 67.36$ km/sec/Mpc, $\Omega_{m}=0.315$, $\Omega_{\Lambda}=0.695$, which are the same as that used in L22 and based on Planck 2020 cosmological parameters~\cite{Planck20}. For super-luminal propagation, one needs to multiply $\Delta t_{\mathrm{LIV}}$ in Eq.~\ref{eq:spectrallag} by -1.

\section{Bayesian Model Comparison}
\label{sec:bayes}
We evaluate the significance of any LIV using Bayesian Model Comparison.  We provide a very brief prelude to Bayesian model comparison, and more details can be found  in recent reviews~\cite{Trotta,Weller,Sanjib,Krishak4}.
To evaluate the significance of a model ($M_2$) as compared to another model ($M_1$), one usually calculates the Bayes factor ($B_{21}$) given by:
\begin{equation}
B_{21}=    \frac{\int P(D|M_2, \theta_2)P(\theta_2|M_2) \, d\theta_2}{\int P(D|M_1, \theta_1)P(\theta_1|M_1) \, d\theta_1} ,  \label{eq:BF}
\end{equation}
where $P(D|M_2,\theta_2)$ is the likelihood for the model $M_2$ given the data $D$ and $P(\theta_2|M_2)$ denotes the prior on the parameter vector $\theta_2$ of the model $M_2$.
The denominator in Eq.~\ref{eq:BF} denotes the same for model $M_1$. If $B_{21}$ is greater than one, then $M_2$ is preferred over $M_1$ and vice-versa. The significance can be qualitatively assessed using the Jeffreys' scale~\cite{Trotta}. 

In the present paper, the model $M_1$ corresponds to the hypothesis, where the spectral lags are produced    only by intrinsic astrophysical emission (Eq.~\ref{eq:int} or Eq.~\ref{eq:intwei}), whereas $M_2$ corresponds to the lags being described by Eq~\ref{eq:totaldelta}, consisting of both intrinsic and LIV delays. To calculate the Bayes factor, we need a model for the likelihood ($\mathcal{L}$), which we define as:
\begin{equation}
     \mathcal{L}=\prod_{i=1}^N \frac{1}{\sigma_{t} \sqrt{2\pi}} \exp \left\{-\frac{[\Delta t _i-f(\Delta E_i,\theta)]^2}{2\sigma_{tot}^2}\right\},
     \label{eq:likelihood}
  \end{equation}
where $N$ is the total number of spectral lags per GRB; $\Delta t_i$ denotes the observed spectral lag data, and  $\sigma_{tot}$ denotes the observed uncertainty, \rthis{which takes into account the uncertainties in the observed spectral lag ($\sigma_t$) as well as the uncertainty in energy ($\sigma_E$)} according to:
\begin{equation}
\sigma_{tot}^2 = \sigma_t^2 + \Big(\frac{\partial f}{\partial E}\Big)^2\sigma_E^2.
\label{eq:xerr}
\end{equation}
In Eq.~\ref{eq:likelihood} and Eq.~\ref{eq:xerr}, $f$ corresponds to  the particular model  being tested, which could either be the null hypothesis of only astrophysical emission or a combination of either of the two LIV models along with the  null hypothesis.

Finally, to  evaluate Eq.~\ref{eq:BF}, we need the priors for the three models. We have used uniform priors for all the intrinsic parameters, and log-uniform priors on $E_{QG}$. The prior ranges for all these parameters for both the LIV and the two intrinsic models considered can be found in Table~\ref{LIU:priortable}.  

\begin{table}[!ht]
    \centering
    \setlength\extrarowheight{-3pt}
    \begin{tabular}{ccc}
        Quantity & Min & Max \\ \hline
        $E_b$(keV) & 0 & 5000\\ 
        $\alpha_1$ & -3 & 10 \\ 
        $\alpha_2$ & -10 & 3 \\ 
        $\mu$ & 0 & 3 \\ 
        $\zeta$ & 0 & 4 \\ 
        $ \log_{10}(E_{QG_1}/GeV)$ & 0 & 20 \\ 
        $ \log_{10}(E_{QG_2}/GeV)$ & 0 & 15 \\ \hline
        $\alpha$ & -2 & 1 \\ 
        $\tau$ & -15 & 10 \\ 
        $ \log_{10}(E_{QG}/GeV)$ & 0 & 20 \\ \hline
    \end{tabular}
    \caption{Summary of priors used for the two intrinsic models (Eq.~\ref{eq:int} and Eq.~\ref{eq:intwei}) as well as the LIV models defined in Eq.~\ref{eq:spectrallag}.}
    \label{LIU:priortable}
\end{table}

\section{Results}
\label{sec:results}
\subsection{Sub-luminal LIV}
We now calculate the Bayes factors assuming  that the spectral lags can be described using a superposition of intrinsic emission  along with a model of LIV compared to only intrinsic emission.  To evaluate the Bayesian evidence we use the {\tt Dynesty} Nested Sampler~\cite{dynesty}. 
Along with the Bayes factor, we also check  the 
efficacy of our fits based on the reduced $\chi^2$, where the reduced $\chi^2$ is equal to the $\chi^2$ divided by the total degrees of freedom.
These values of the Bayes factor and reduced $\chi^2$ for all the 32 GRBs considered can be found in Table~\ref{LIU:bf} for the SPBL model as the null hypothesis.  For linear LIV, we find \rthis{five} GRBs with Bayes factor $> 100$. These are \rthis{GRB 210204A}, GRB 190114C,  \rthis{GRB 160625B, GRB 180703A}, and   GRB 130925A. For quadratic LIV also, \rthis{five GRBs have Bayes factors $> 100$, viz. GRB 210204A, GRB 190114C, GRB 160625B, GRB 131108A, GRB 130925A.}

The corresponding results for the more simple power law model in  Eq.~\ref{eq:intwei} can be found in Table~\ref{AWL:bf}. In contrast to the SBPL intrinsic model, now we find \rthis{10}  GRBs with Bayes factors $> 100$ for linear LIV model. On the other hand for the quadratic model of LIV, we find \rthis{nine} GRBs with Bayes factor $> 100$.

We also find that the reduced $\chi^2$ for the null hypothesis is larger while using Eq.~\ref{eq:intwei}, as compared to the SBPL parameterization. In order to validate this with Bayesian model comparison, we compare the Bayesian evidence for both these intrinsic models. We again use the same priors for both these models as before. The Bayes factors for the SBPL model compared to Eq.~\ref{eq:intwei} can be found in Table~\ref{tab:int} for all the 32 GRBs. We find  that \rthis{16} GRBs show decisive evidence in favor of the SBPL model. The more simple power law model (Eq.~\ref{eq:intwei}) is favored compared to the SBPL for only  \rthis{three} GRBs. This agrees with the conclusions in L22, who pointed out that the SBPL model is more accurate than the simple power model used  before in literature.

Finally, it is  instructive to compare the statistical significance of LIV for GRB~190114C and GRB~1606025B with previous works on the subject, which have used Eq.~\ref{eq:intwei} for the null hypothesis. For  GRB 190114C, the natural log of Bayes factor was  approximately 175 for linear and quadratic LIV~\cite{Du}. We find the same to be about \rthis{41 and 35} for linear and quadratic LIV, respectively using the same intrinsic model. However, when we use the SBPL as the null hypothesis, the corresponding natural log of the Bayes factors  reduce to \rthis{6.85 and 6.1},  respectively. \rthis{However, these Bayes factors still correspond to decisive evidence.}

For GRB 1606025B, frequentist, information theory,  and  Bayesian  model comparison techniques have been used to evaluate the statistical significance of LIV~\cite{Ganguly,Gunapati}. ~\citet{Gunapati} reported the natural log of Bayes factor of 16 and 20, for linear and quadratic LIV using the intrinsic model in Eq.~\ref{eq:intwei}.  The corresponding values for the natural lag of the Bayes factors, which we get compared to the same null hypothesis  are comparable with values of \rthis{20.6 and 12.5} for linear and quadratic LIV, respectively. These Bayes factors correspond to decisive evidence for LIV.
\rthis{When we consider the SBPL model as the null hypothesis, the corresponding nature log of the Bayes factors increase to 20.5 and 44.3.} This underscores the importance of correctly modelling the intrinsic emission, while drawing conclusions about LIV from spectral lag data. 


\begin{table}[H]
\footnotesize
\centering
\setlength\extrarowheight{-3pt}
\begin{tabular}{|c|c|c|cc|cc|cc|} 
\hline
\multirow{2}{*}{\textbf{GRB}}        & \multirow{2}{*}{\textbf{E0 (keV)}}        & \multirow{2}{*}{\textbf{Redshift}} & 
\multicolumn{2}{|c|}{\textbf{Null}} & \multicolumn{2}{|c|}{\textbf{Null + Linear LIV}} & \multicolumn{2}{|c|}{\textbf{Null + Quadratic LIV}}\\
&&& \textbf{$\ln$(BF)} & \textbf{Reduced $\chi^2$} & \textbf{$\ln$(BF)} & \textbf{Reduced $\chi^2$} & \textbf{$ \ln$(BF)} & \textbf{Reduced $\chi^2$}\\
\hline
GRB 210619B & 10.00 & 1.94 & 0.00 & 2.84 & -7.50 & 3.13 & 0.55 & 2.74 \\
GRB 210610B & 30.00 & 1.13 & 0.00 & 1.14 & -1.24 & 0.91 & -0.53 & 1.02 \\
GRB 210204A & 10.00 & 0.88 & 0.00 & 5.06 & 10.51 & 3.85 & 7.39 & 3.97 \\
GRB 201216C & 15.00 & 1.10 & 0.00 & 1.07 & -1.23 & 1.08 & -0.58 & 1.02 \\
GRB 200829A & 25.00 & 1.25 & 0.00 & 2.01 & -0.04 & 3.47 & 2.19 & 2.13 \\
GRB 200613A & 30.00 & 1.22 & 0.00 & 0.75 & -0.92 & 0.79 & -0.26 & 0.77 \\
GRB 190114C & 10.00 & 0.42 & 0.00 & 3.51 & 6.85 & 1.48 & 6.10 & 1.18 \\
GRB 180720B & 25.00 & 0.65 & 0.00 & 0.90 & -1.66 & 0.97 & -0.82 & 1.13 \\
GRB 180703A & 20.00 & 0.67 & 0.00 & 8.95 & 6.53 & 1.84 & -0.55 & 9.64 \\
GRB 171010A & 10.00 & 0.33 & 0.00 & 0.63 & -0.63 & 0.69 & 0.35 & 0.58 \\
GRB 160625B & 10.00 & 1.41 & 0.00 & 6.54 & 20.48 & 5.44 & 44.34 & 6.12 \\
GRB 160509A & 10.00 & 1.17 & 0.00 & 0.72 & -1.05 & 0.78 & -0.42 & 1.27 \\
GRB 150821A & 10.00 & 0.76 & 0.00 & 0.80 & -0.99 & 0.67 & -0.40 & 0.86 \\
GRB 150514A & 20.00 & 0.81 & 0.00 & 1.23 & -1.01 & 1.16 & -0.26 & 1.51 \\
GRB 150403A & 35.00 & 2.06 & 0.00 & 1.51 & -0.91 & 0.97 & -0.13 & 0.87 \\
GRB 150314A & 20.00 & 1.76 & 0.00 & 3.89 & -1.98 & 3.97 & -2.34 & 3.49 \\
GRB 141028A & 40.00 & 2.33 & 0.00 & 0.89 & -1.07 & 0.68 & -0.26 & 1.17 \\
GRB 140508A & 10.00 & 1.03 & 0.00 & 0.91 & -0.88 & 0.79 & -0.29 & 1.97 \\
GRB 140206A & 20.00 & 2.73 & 0.00 & 2.09 & -1.33 & 2.20 & -1.23 & 1.98 \\
GRB 131231A & 10.00 & 0.64 & 0.00 & 1.77 & -1.60 & 1.81 & -0.36 & 1.76 \\
GRB 131108A & 20.00 & 2.40 & 0.00 & 4.41 & 1.11 & 3.39 & 5.18 & 3.71 \\
GRB 130925A & 10.00 & 0.35 & 0.00 & 3.62 & 4.82 & 2.09 & 6.64 & 1.83 \\
GRB 130518A & 10.00 & 2.49 & 0.00 & 1.92 & 0.33 & 1.66 & 0.03 & 2.09 \\
GRB 130427A & 10.00 & 0.34 & 0.00 & 0.38 & 0.76 & 0.46 & -0.11 & 1.00 \\
GRB 120119A & 25.00 & 1.73 & 0.00 & 1.25 & -0.92 & 1.22 & -0.71 & 0.78 \\
GRB 100728A & 40.00 & 1.57 & 0.00 & 2.59 & -1.64 & 2.77 & -0.96 & 2.50 \\
GRB 091003A & 10.00 & 0.90 & 0.00 & 2.54 & -0.56 & 2.62 & -0.18 & 3.22 \\
GRB 090926A & 10.00 & 2.11 & 0.00 & 0.34 & -1.36 & 0.29 & -0.49 & 0.34 \\
GRB 090618 & 10.00 & 0.54 & 0.00 & 0.33 & -1.37 & 0.44 & -0.14 & 0.65 \\
GRB 090328 & 30.00 & 0.74 & 0.00 & 5.38 & -0.99 & 5.25 & 0.06 & 5.27 \\
GRB 081221 & 10.00 & 2.26 & 0.00 & 0.48 & -1.37 & 0.87 & -0.68 & 0.98 \\
GRB 080916C & 10.00 & 4.35 & 0.00 & 1.19 & -1.37 & 1.19 & -0.61 & 1.24 \\
\hhline{#=========#}
\end{tabular}
\caption{\label{LIU:bf}Bayes factors and reduced $\chi^2$ for linear and quadratic \rthis{sub-luminal} LIV models along with the SBPL intrinsic emission model  considered in Eq.~\ref{eq:int}, compared to only intrinsic emission model for every GRB considered in this work. We also list the corresponding redshift and lower value for the energy bin  ($E_0$) used to calculate the spectral lag. For linear LIV, we find \rthis{five} GRBs with Bayes factors $> 100$ (or $\ln~\rm{BF}> 4.6$), viz.  GRB 210204A, GRB 190114C, GRB 180703A, GRB 160625B, GRB 130925A. \rthis{For quadratic LIV, we find five GRBs with a Bayes factor $> 100$, viz. GRB 210204A, GRB 190114C, GRB 160625B, GRB 131108A, GRB 130925A.}
}

\end{table}


\begin{table}[H]
\footnotesize
\centering
\setlength\extrarowheight{-3pt}
\begin{tabular}{|c|c|c|cc|cc|cc|} 
\hline
\multirow{2}{*}{\textbf{GRB}}        & \multirow{2}{*}{\textbf{E0 (keV)}}        & \multirow{2}{*}{\textbf{Redshift}} & 
\multicolumn{2}{|c|}{\textbf{Null}} & \multicolumn{2}{|c|}{\textbf{Null + Linear LIV}} & \multicolumn{2}{|c|}{\textbf{Null + Quadratic LIV}}\\
&&& \textbf{$\ln$(BF)} & \textbf{Reduced $\chi^2$} & \textbf{$\ln$(BF)} & \textbf{Reduced $\chi^2$} & \textbf{$ \ln$(BF)} & \textbf{Reduced $\chi^2$}\\
\hline
GRB 210619B & 10.00 & 1.94 & 0.00 & 6.48 & 15.64 & 4.56 & 13.33 & 4.93 \\
GRB 210610B & 30.00 & 1.13 & 0.00 & 0.99 & -0.76 & 1.02 & 0.39 & 0.61 \\
GRB 210204A & 10.00 & 0.88 & 0.00 & 3.56 & 3.84 & 3.38 & 4.67 & 3.10 \\
GRB 201216C & 15.00 & 1.10 & 0.00 & 1.04 & -0.72 & 1.13 & -0.05 & 1.14 \\
GRB 200829A & 25.00 & 1.25 & 0.00 & 4.71 & 71.15 & 4.36 & 73.43 & 4.06 \\
GRB 200613A & 30.00 & 1.22 & 0.00 & 0.80 & -1.05 & 0.69 & -0.24 & 0.65 \\
GRB 190114C & 10.00 & 0.42 & 0.00 & 2.01 & 40.98 & 2.11 & 35.41 & 2.24 \\
GRB 180720B & 25.00 & 0.65 & 0.00 & 0.96 & -1.06 & 0.72 & -0.08 & 0.91 \\
GRB 180703A & 20.00 & 0.67 & 0.00 & 3.14 & 8.22 & 2.21 & 0.32 & 2.88 \\
GRB 171010A & 10.00 & 0.33 & 0.00 & 0.48 & -0.98 & 0.43 & -0.14 & 0.42 \\
GRB 160625B & 10.00 & 1.41 & 0.00 & 3.05 & 20.60 & 2.11 & 12.52 & 2.81 \\
GRB 160509A & 10.00 & 1.17 & 0.00 & 1.07 & -1.12 & 0.80 & -0.16 & 0.98 \\
GRB 150821A & 10.00 & 0.76 & 0.00 & 0.83 & -1.23 & 1.01 & -0.39 & 0.96 \\
GRB 150514A & 20.00 & 0.81 & 0.00 & 0.90 & -0.92 & 0.69 & -0.18 & 1.02 \\
GRB 150403A & 35.00 & 2.06 & 0.00 & 0.74 & -1.09 & 1.08 & -0.18 & 0.80 \\
GRB 150314A & 20.00 & 1.76 & 0.00 & 6.21 & 2.43 & 3.62 & 3.46 & 5.50 \\
GRB 141028A & 40.00 & 2.33 & 0.00 & 0.80 & -1.22 & 0.86 & -0.32 & 0.99 \\
GRB 140508A & 10.00 & 1.03 & 0.00 & 0.83 & -1.05 & 1.06 & -0.26 & 0.90 \\
GRB 140206A & 20.00 & 2.73 & 0.00 & 2.40 & 19.56 & 1.09 & 19.47 & 2.46 \\
GRB 131231A & 10.00 & 0.64 & 0.00 & 1.64 & -0.20 & 1.25 & 2.20 & 0.93 \\
GRB 131108A & 20.00 & 2.40 & 0.00 & 4.49 & 14.77 & 2.74 & 1.65 & 5.42 \\
GRB 130925A & 10.00 & 0.35 & 0.00 & 0.67 & -1.13 & 0.72 & -0.06 & 0.62 \\
GRB 130518A & 10.00 & 2.49 & 0.00 & 3.89 & 26.96 & 2.27 & 27.97 & 1.67 \\
GRB 130427A & 10.00 & 0.34 & 0.00 & 5.86 & 34.65 & 2.72 & 25.30 & 3.50 \\
GRB 120119A & 25.00 & 1.73 & 0.00 & 1.65 & 0.39 & 1.96 & 0.99 & 1.79 \\
GRB 100728A & 40.00 & 1.57 & 0.00 & 2.32 & -0.76 & 2.36 & 0.15 & 2.07 \\
GRB 091003A & 10.00 & 0.90 & 0.00 & 4.06 & 1.73 & 3.53 & 1.82 & 4.47 \\
GRB 090926A & 10.00 & 2.11 & 0.00 & 1.05 & 0.76 & 0.45 & 2.34 & 0.29 \\
GRB 090618 & 10.00 & 0.54 & 0.00 & 0.61 & -1.16 & 0.85 & -0.43 & 0.88 \\
GRB 090328 & 30.00 & 0.74 & 0.00 & 6.86 & 13.33 & 2.22 & 5.85 & 5.58 \\
GRB 081221 & 10.00 & 2.26 & 0.00 & 1.97 & -0.53 & 1.96 & 0.49 & 1.42 \\
GRB 080916C & 10.00 & 4.35 & 0.00 & 2.18 & 1.33 & 1.22 & 0.14 & 2.43 \\
\hhline{#=========#}
\end{tabular}
\caption{\label{AWL:bf}  Bayes factors for linear and quadratic LIV \rthis{sub-luminal} models along with the simple power law intrinsic emission model considered in Eq.~\ref{eq:intwei}, compared to only intrinsic emission. See Table~\ref{LIU:bf} for explanation of the various columns. We find 10 and 9 GRBs with Bayes factors $> 100$ for linear and quadratic LIV models, respectively.}
\end{table}

\begin{table}[H]
\footnotesize
\centering
\setlength\extrarowheight{-3pt}
    \centering
    \begin{tabular}{|c|c|}
    \hline
\textbf{GRB} & $\ln \rm{[BF_{SBPL} (Eq.~\ref{eq:int})]}$ - $\ln~\rm{[BF~(Eq.~\ref{eq:intwei})]}$ \\ 
\hline
GRB 210619B & 25.73 \\
GRB 210610B & 20.42 \\
GRB 210204A & -8.39 \\
GRB 201216C & -2.82 \\
GRB 200829A & 77.98 \\
GRB 200613A & -3.17 \\
GRB 190114C & 33.07 \\
GRB 180720B & 2.25 \\
GRB 180703A & 21.14 \\
GRB 171010A & 13.50 \\
GRB 160625B & -75.70 \\
GRB 160509A & -1.67 \\
GRB 150821A & 1.10 \\
GRB 150514A & -3.25 \\
GRB 150403A & -3.45 \\
GRB 150314A & 19.71 \\
GRB 141028A & -2.41 \\
GRB 140508A & -3.40 \\
GRB 140206A & 21.85 \\
GRB 131231A & 17.24 \\
GRB 131108A & 20.01 \\
GRB 130925A & -5.04 \\
GRB 130518A & 27.67 \\
GRB 130427A & 57.02 \\
GRB 120119A & 0.34 \\
GRB 100728A & 9.56 \\
GRB 091003A & 8.78 \\
GRB 090926A & 3.25 \\
GRB 090618 & 3.08 \\
GRB 090328 & 13.48 \\
GRB 081221 & 5.11 \\
GRB 080916C & 4.20 \\
\hhline{#==#}
\end{tabular}
    \caption{\label{tab:int} Bayes factor (in natural log) for the intrinsic model specified by the SBPL model compared to the intrinsic model in Eq.~\ref{eq:intwei}. For most  GRBs  \rthis{(16)}, the Bayes model comparison decisively  favors the SBPL parameterization compared to that in Eq.~\ref{eq:intwei}. }
\end{table}

\subsection{Super-luminal LIV}
\label{sec:superluminal}
We now carry out a similar search for LIV assuming super-luminal propagation. For super-luminal  propagation, $\Delta t_{LIV}$ in Eq.~\ref{eq:spectrallag} needs to be multiplied by -1.  For the  intrinsic emission model, we use both the SBPL model similar to L22 as well as the simple power model (Eq.~\ref{eq:intwei}).
Our results from this analysis for all the GRBs can be found in Table~\ref{tab:SLPL} (for the simple power law intrinsic model)  and in Table~\ref{tab:superluminal} (for the SBPL intrinsic model). 

For the simple power intrinsic model, we find   \rthis{four GRBs (viz. GRB 200829A, GRB 180703A, GRB 131108A, GRB 090328) with  Bayes factors $> 100$ for linear LIV. For quadratic LIV, only two of them (GRB 180703A, GRB 131108A) show a Bayes factor $> 100$.}
When we consider the SBPL model, 
only \rthis{one  GRBs (GRB 160625B)  shows  Jeffreys scale based ``decisive evidence''  with Bayes factors $>100$ for both linear and quadratic LIV.}  Therefore,  there are significant differences in the results for sub-luminal and super-luminal propagation.


\begin{table}[H]
\footnotesize
\centering
\setlength\extrarowheight{-3pt}
\begin{tabular}{|c|c|c|cc|cc|cc|} 
\hline
\multirow{2}{*}{\textbf{GRB}}        & \multirow{2}{*}{\textbf{E0 (keV)}}        & \multirow{2}{*}{\textbf{Redshift}} & 
\multicolumn{2}{|c|}{\textbf{Null}} & \multicolumn{2}{|c|}{\textbf{Null + Linear LIV}} & \multicolumn{2}{|c|}{\textbf{Null + Quadratic LIV}}\\
&&& \textbf{$\ln$(BF)} & \textbf{Reduced $\chi^2$} & \textbf{$\ln$(BF)} & \textbf{Reduced $\chi^2$} & \textbf{$ \ln$(BF)} & \textbf{Reduced $\chi^2$}\\
\hline
GRB 210619B & 10.00 & 1.94 & 0.00 & 6.33 & -2.78 & 6.84 & -1.06 & 6.89 \\
GRB 210610B & 30.00 & 1.13 & 0.00 & 1.16 & -1.21 & 0.57 & -0.27 & 1.31 \\
GRB 210204A & 10.00 & 0.88 & 0.00 & 2.62 & -1.66 & 2.72 & -0.78 & 3.00 \\
GRB 201216C & 15.00 & 1.10 & 0.00 & 1.04 & -1.57 & 1.08 & -0.46 & 1.21 \\
GRB 200829A & 25.00 & 1.25 & 0.00 & 4.49 & 12.53 & 2.31 & 0.09 & 5.12 \\
GRB 200613A & 30.00 & 1.22 & 0.00 & 0.75 & -1.41 & 0.71 & -0.37 & 0.79 \\
GRB 190114C & 10.00 & 0.42 & 0.00 & 2.75 & -1.36 & 2.99 & -0.18 & 2.20 \\
GRB 180720B & 25.00 & 0.65 & 0.00 & 0.84 & -1.72 & 1.13 & -0.50 & 1.13 \\
GRB 180703A & 20.00 & 0.67 & 0.00 & 3.00 & 30.34 & 1.63 & 9.13 & 2.18 \\
GRB 171010A & 10.00 & 0.33 & 0.00 & 0.42 & -0.97 & 0.72 & -0.26 & 0.38 \\
GRB 160625B & 10.00 & 1.41 & 0.00 & 3.55 & -2.05 & 3.01 & -0.25 & 3.08 \\
GRB 160509A & 10.00 & 1.17 & 0.00 & 0.84 & -1.28 & 0.86 & -0.05 & 0.97 \\
GRB 150821A & 10.00 & 0.76 & 0.00 & 0.76 & -1.05 & 0.81 & -0.40 & 1.01 \\
GRB 150514A & 20.00 & 0.81 & 0.00 & 1.26 & -1.32 & 0.95 & -0.45 & 1.03 \\
GRB 150403A & 35.00 & 2.06 & 0.00 & 0.73 & -1.53 & 0.85 & -0.43 & 0.88 \\
GRB 150314A & 20.00 & 1.76 & 0.00 & 6.19 & -1.43 & 6.53 & -0.49 & 6.58 \\
GRB 141028A & 40.00 & 2.33 & 0.00 & 0.84 & -1.41 & 0.86 & -0.52 & 0.86 \\
GRB 140508A & 10.00 & 1.03 & 0.00 & 0.94 & -1.55 & 0.93 & -0.58 & 0.84 \\
GRB 140206A & 20.00 & 2.73 & 0.00 & 2.32 & -0.89 & 4.13 & 0.08 & 2.23 \\
GRB 131231A & 10.00 & 0.64 & 0.00 & 1.62 & -1.31 & 0.95 & -0.52 & 1.74 \\
GRB 131108A & 20.00 & 2.40 & 0.00 & 4.54 & 9.35 & 1.99 & 5.53 & 6.03 \\
GRB 130925A & 10.00 & 0.35 & 0.00 & 0.52 & -1.08 & 0.89 & -0.37 & 0.56 \\
GRB 130518A & 10.00 & 2.49 & 0.00 & 3.25 & -1.79 & 3.79 & -0.99 & 2.58 \\
GRB 130427A & 10.00 & 0.34 & 0.00 & 5.97 & -1.91 & 2.16 & -0.65 & 6.27 \\
GRB 120119A & 25.00 & 1.73 & 0.00 & 1.61 & -1.58 & 1.77 & -0.63 & 1.75 \\
GRB 100728A & 40.00 & 1.57 & 0.00 & 2.42 & -1.27 & 2.61 & -0.17 & 2.48 \\
GRB 091003A & 10.00 & 0.90 & 0.00 & 4.15 & -1.73 & 4.35 & -0.29 & 4.59 \\
GRB 090926A & 10.00 & 2.11 & 0.00 & 1.04 & -1.53 & 1.26 & -0.12 & 1.17 \\
GRB 090618 & 10.00 & 0.54 & 0.00 & 0.63 & -1.16 & 0.94 & -0.45 & 0.80 \\
GRB 090328 & 30.00 & 0.74 & 0.00 & 7.03 & 6.35 & 1.92 & -0.54 & 7.60 \\
GRB 081221 & 10.00 & 2.26 & 0.00 & 1.90 & -1.43 & 2.03 & -0.30 & 2.07 \\
GRB 080916C & 10.00 & 4.35 & 0.00 & 2.12 & -1.50 & 2.35 & -0.22 & 2.19 \\
\hline
\end{tabular}
\caption{\label{tab:SLPL} Results of searches for LIV  using super-luminal propagation and a simple power law model for intrinsic emission (Eq.~\ref{eq:intwei}) For explanation of all columns, see  Table~\ref{LIU:bf}. We find that \rthis{four GRBs (viz. GRB 200829A, GRB 180703A, GRB 131108A, GRB 090328) show a Bayes factor $> 100$ for linear LIV. For quadratic LIV, only two of them (GRB 180703A, GRB 131108A) show a Bayes factor $> 100$.}
}
\end{table}

\begin{table}[H]
\footnotesize
\centering
\setlength\extrarowheight{-3pt}
\begin{tabular}{|c|c|c|cc|cc|cc|} 
\hline
\multirow{2}{*}{\textbf{GRB}}        & \multirow{2}{*}{\textbf{E0 (keV)}}        & \multirow{2}{*}{\textbf{Redshift}} & 
\multicolumn{2}{|c|}{\textbf{Null}} & \multicolumn{2}{|c|}{\textbf{Null + Linear LIV}} & \multicolumn{2}{|c|}{\textbf{Null + Quadratic LIV}}\\
&&& \textbf{$\ln$(BF)} & \textbf{Reduced $\chi^2$} & \textbf{$\ln$(BF)} & \textbf{Reduced $\chi^2$} & \textbf{$ \ln$(BF)} & \textbf{Reduced $\chi^2$}\\
\hline
GRB 210619B & 10.00 & 1.94 & 0.00 & 3.30 & -1.24 & 2.47 & -0.55 & 2.14 \\
GRB 210610B & 30.00 & 1.13 & 0.00 & 1.12 & -1.70 & 1.01 & -1.17 & 0.65 \\
GRB 210204A & 10.00 & 0.88 & 0.00 & 5.30 & -3.51 & 5.92 & -5.10 & 6.43 \\
GRB 201216C & 15.00 & 1.10 & 0.00 & 1.09 & -0.97 & 1.10 & -0.21 & 1.05 \\
GRB 200829A & 25.00 & 1.25 & 0.00 & 1.95 & -3.35 & 1.91 & -4.00 & 2.19 \\
GRB 200613A & 30.00 & 1.22 & 0.00 & 1.13 & -1.38 & 1.57 & -0.57 & 0.91 \\
GRB 190114C & 10.00 & 0.42 & 0.00 & 3.46 & -2.18 & 3.52 & -0.50 & 3.86 \\
GRB 180720B & 25.00 & 0.65 & 0.00 & 1.00 & -1.07 & 1.29 & -0.58 & 1.18 \\
GRB 180703A & 20.00 & 0.67 & 0.00 & 9.09 & -1.58 & 10.06 & -0.56 & 10.19 \\
GRB 171010A & 10.00 & 0.33 & 0.00 & 0.43 & -1.71 & 0.52 & -0.79 & 0.60 \\
GRB 160625B & 10.00 & 1.41 & 0.00 & 7.92 & 8.76 & 8.75 & 64.00 & 4.81 \\
GRB 160509A & 10.00 & 1.17 & 0.00 & 0.73 & -1.52 & 0.94 & -0.72 & 1.39 \\
GRB 150821A & 10.00 & 0.76 & 0.00 & 1.04 & -1.48 & 0.90 & -0.02 & 0.92 \\
GRB 150514A & 20.00 & 0.81 & 0.00 & 1.11 & -1.07 & 1.30 & -0.49 & 1.22 \\
GRB 150403A & 35.00 & 2.06 & 0.00 & 0.89 & -1.42 & 1.29 & -0.59 & 1.08 \\
GRB 150314A & 20.00 & 1.76 & 0.00 & 3.45 & 0.27 & 3.71 & 2.31 & 3.01 \\
GRB 141028A & 40.00 & 2.33 & 0.00 & 1.21 & -1.34 & 1.27 & -0.38 & 0.97 \\
GRB 140508A & 10.00 & 1.03 & 0.00 & 1.49 & -1.72 & 0.86 & -0.87 & 1.35 \\
GRB 140206A & 20.00 & 2.73 & 0.00 & 1.40 & -1.48 & 2.22 & -1.51 & 1.57 \\
GRB 131231A & 10.00 & 0.64 & 0.00 & 1.62 & -3.43 & 2.25 & -1.00 & 2.04 \\
GRB 131108A & 20.00 & 2.40 & 0.00 & 4.79 & -1.51 & 5.14 & -0.82 & 4.76 \\
GRB 130925A & 10.00 & 0.35 & 0.00 & 3.22 & -1.66 & 3.51 & 0.01 & 3.29 \\
GRB 130518A & 10.00 & 2.49 & 0.00 & 2.25 & -1.56 & 2.22 & -0.74 & 2.19 \\
GRB 130427A & 10.00 & 0.34 & 0.00 & 1.01 & 2.81 & 0.40 & 0.47 & 0.46 \\
GRB 120119A & 25.00 & 1.73 & 0.00 & 1.20 & -1.50 & 1.60 & -0.55 & 1.58 \\
GRB 100728A & 40.00 & 1.57 & 0.00 & 2.39 & -1.02 & 2.57 & -0.11 & 2.60 \\
GRB 091003A & 10.00 & 0.90 & 0.00 & 2.60 & -1.87 & 3.18 & -0.32 & 2.92 \\
GRB 090926A & 10.00 & 2.11 & 0.00 & 0.37 & -0.63 & 0.74 & 0.03 & 0.35 \\
GRB 090618 & 10.00 & 0.54 & 0.00 & 0.42 & -1.66 & 0.68 & -0.24 & 0.95 \\
GRB 090328 & 30.00 & 0.74 & 0.00 & 4.58 & -0.83 & 4.86 & -0.68 & 4.89 \\
GRB 081221 & 10.00 & 2.26 & 0.00 & 0.73 & -1.25 & 0.74 & -0.71 & 0.55 \\
GRB 080916C & 10.00 & 4.35 & 0.00 & 0.98 & -1.46 & 1.66 & -0.67 & 1.44 \\
\hline
\end{tabular}
\caption{\label{tab:superluminal} Results for  super-luminal LIV using the SBPL model for intrinsic emission as in L22. For explanation of all columns, see  Table~\ref{LIU:bf}. For both linear and quadratic LIV, only GRB 160625B show a Bayes factor $> 100$. 
}
\end{table}

\section{Conclusions}
\label{sec:conclusions}
In a recent work, L22 carried out a comprehensive study of the spectral lags of 32 long GRBs detected by Fermi-GBM, which had a transition from positive to negative lags. They fit the intrinsic lags using an empirical SBPL (Eq.~\ref{eq:int}). L22 used this data to constrain LIV and obtained constraints on $E_{QG} \geq 1.5 \times 10^{14}$ GeV 
and $E_{QG} \geq 8 \times 10^{5}$ GeV.

In this work, we extended the original analysis in L22 by evaluating the statistical significance based on Bayesian model selection, that the spectral lags can be adequately modelled by a mixture of intrinsic emission and LIV-induced compared to only intrinsic emission. We also did a search for super-luminal LIV. For the intrinsic emission, we consider two models.  One of them is the SBPL model considered in L22. The second intrinsic model we consider  is the simple   power law  model (cf. Eq.~\ref{eq:intwei}) first used in ~\cite{Wei},  and which was used in some of our past works~\cite{Ganguly,Agrawal_2021,Desai23}. \rthis{We also incorporated the uncertainties in the energies for our analysis}

Our results for the Bayes factor and reduced $\chi^2$
for sub-luminal LIV can be found for the  Table~\ref{LIU:bf} and Table~\ref{AWL:bf} for the SBPL model and   power law (Eq.~\ref{eq:intwei}), respectively.  The corresponding results for super-luminal LIV can be found in Table~\ref{tab:SLPL} and Table~\ref{tab:superluminal} for the power law and SBPL intrinsic models, respectively.
To evaluate the relative efficacy of the two models of intrinsic emission, we also calculate the Bayes factor for the SBPL model compared to that in Eq.~\ref{eq:intwei}. These Bayes factors can be found for all the GRBs in Tab.~\ref{tab:int}. 
Our conclusions are as follows:

\begin{itemize}
\item For sub-luminal propagation, we find \rthis{five} GRBs (GRB 210204A, GRB 190114C, GRB 130925A, GRB 180703A, and GRB 130925A) with Bayes factor $> 100$  (corresponding to Jeffreys scale based ``decisive evidence'') for a model consisting of a  superposition of SBPL  + linear LIV compared to only the SBPL model.   \rthis{For quadratic LIV, except GRB 180703A, all the aforementioned GRBs and GRB 131108A show ``Jeffreys scale'' based decisive evidence for LIV.}

\item For sub-luminal propagation, When we replace the SBPL model with Eq.~\ref{eq:intwei} as the  null hypothesis, we find 10 and 9 GRBs with  Jeffreys scale based ``decisive evidence'' for linear and quadratic LIV, respectively.

\item For most GRBs, Bayesian model comparison decisively favors the SBPL model compared to the intrinsic model used in Eq.~\ref{eq:intwei}. This is in accord with the conclusions in L22.


\item For super-luminal propagation using  the simple power intrinsic model, we find that \rthis{four GRBs (viz. GRB 200829A, GRB 180703A, GRB 131108A, GRB 090328) show  Jeffreys scale  based ``decisive evidence'' for linear LIV.  For quadratic LIV,  two GRBs (GRB 180703A, GRB 131108A) show Jeffreys scale  based ``decisive evidence''.}

\item For super-luminal propagation, when we consider the SBPL model we find that for  quadratic  LIV, only \rthis{one GRB  (GRB 160625B)  shows Bayes factors $> 100$, for  both linear and quadratic LIV.} Therefore the results for super-luminal LIV differ considerably compared to those for sub-luminal LIV.

We should however caution that the large Bayes factors which we obtain (assuming the spectral lags are caused by  a model of LIV and intrinsic emission compared to only intrinsic emission)  should not be construed as a proof of LIV.  One reason is that the reduced $\chi^2$ is much greater than one for most of these LIV models. The bulk of experimental tests, both within and beyond astrophysical contexts, have thus far validated the predictions of Special  Relativity within the explored realms of energy and precision~\cite{Will05}. Noteworthy studies, including those involving high-energy cosmic rays, gamma-ray bursts, and neutrino oscillations, have imposed rigorous limits on potential deviations from Lorentz invariance, without furnishing any indications of LIV~\cite{Desairev,WeiWu2,WeiWu2021}. As expounded by ~\cite{Liu_2022} and ~\cite{Wei},
the main applicability  of such fitting methodologies and the alignment (or lack thereof) between data and models of LIV  primarily  serve to confine the limits of the LIV effect and also to examine the efficacy of different intrinsic emission models, rather than confirming the presence of LIV.

Our analysis codes along with supplementary plots showing comparison of the different models on top of data have been uploaded on {\tt github} and can be found in \url{https://github.com/DarkWake9/Project-QG}

\end{itemize}
\section*{Acknowledgements}
We are grateful to Zik Liu and Binbin Zhang for generously sharing the spectral lag data used in L22 with us. \rthis{We also acknowledge the anonymous referee for very useful constructive feedback on our manuscript.}
We acknowledge National Supercomputing Mission (NSM) for providing computing resources of ‘PARAM SEVA’ at IIT, Hyderabad, which is implemented by C-DAC and supported by the Ministry of Electronics and Information Technology (MeitY) and Department of Science and Technology (DST), Government of India. VP also acknowledges partial support from T-641 (DST-ICPS).

\bibliography{main}

\begin{thebibliography}{39}
\expandafter\ifx\csname natexlab\endcsname\relax\def\natexlab#1{#1}\fi
\expandafter\ifx\csname bibnamefont\endcsname\relax
  \def\bibnamefont#1{#1}\fi
\expandafter\ifx\csname bibfnamefont\endcsname\relax
  \def\bibfnamefont#1{#1}\fi
\expandafter\ifx\csname citenamefont\endcsname\relax
  \def\citenamefont#1{#1}\fi
\expandafter\ifx\csname url\endcsname\relax
  \def\url#1{\texttt{#1}}\fi
\expandafter\ifx\csname urlprefix\endcsname\relax\def\urlprefix{URL }\fi
\providecommand{\bibinfo}[2]{#2}
\providecommand{\eprint}[2][]{\url{#2}}

\bibitem[{\citenamefont{{Liu} et~al.}(2022)\citenamefont{{Liu}, {Zhang}, and
  {Meng}}}]{Liu_2022}
\bibinfo{author}{\bibfnamefont{Z.-K.} \bibnamefont{{Liu}}},
  \bibinfo{author}{\bibfnamefont{B.-B.} \bibnamefont{{Zhang}}},
  \bibnamefont{and} \bibinfo{author}{\bibfnamefont{Y.-Z.}
  \bibnamefont{{Meng}}}, \bibinfo{journal}{\apj}
  \textbf{\bibinfo{volume}{935}}, \bibinfo{eid}{79} (\bibinfo{year}{2022}),
  \eprint{2202.09999}.

\bibitem[{\citenamefont{{Amelino-Camelia}
  et~al.}(1998)\citenamefont{{Amelino-Camelia}, {Ellis}, {Mavromatos},
  {Nanopoulos}, and {Sarkar}}}]{GAA98}
\bibinfo{author}{\bibfnamefont{G.}~\bibnamefont{{Amelino-Camelia}}},
  \bibinfo{author}{\bibfnamefont{J.}~\bibnamefont{{Ellis}}},
  \bibinfo{author}{\bibfnamefont{N.~E.} \bibnamefont{{Mavromatos}}},
  \bibinfo{author}{\bibfnamefont{D.~V.} \bibnamefont{{Nanopoulos}}},
  \bibnamefont{and} \bibinfo{author}{\bibfnamefont{S.}~\bibnamefont{{Sarkar}}},
  \bibinfo{journal}{\nat} \textbf{\bibinfo{volume}{395}}, \bibinfo{pages}{525}
  (\bibinfo{year}{1998}).

\bibitem[{\citenamefont{Addazi et~al.}(2022)}]{Addazi21}
\bibinfo{author}{\bibfnamefont{A.}~\bibnamefont{Addazi}} \bibnamefont{et~al.},
  \bibinfo{journal}{Prog. Part. Nucl. Phys.} \textbf{\bibinfo{volume}{125}},
  \bibinfo{pages}{103948} (\bibinfo{year}{2022}), \eprint{2111.05659}.

\bibitem[{\citenamefont{{Kumar} and {Zhang}}(2015)}]{Kumar}
\bibinfo{author}{\bibfnamefont{P.}~\bibnamefont{{Kumar}}} \bibnamefont{and}
  \bibinfo{author}{\bibfnamefont{B.}~\bibnamefont{{Zhang}}},
  \bibinfo{journal}{\physrep} \textbf{\bibinfo{volume}{561}},
  \bibinfo{pages}{1} (\bibinfo{year}{2015}), \eprint{1410.0679}.

\bibitem[{\citenamefont{{Yu} et~al.}(2022)\citenamefont{{Yu}, {Gao}, {Wang},
  and {Zhang}}}]{WuGRBreview}
\bibinfo{author}{\bibfnamefont{Y.-W.} \bibnamefont{{Yu}}},
  \bibinfo{author}{\bibfnamefont{H.}~\bibnamefont{{Gao}}},
  \bibinfo{author}{\bibfnamefont{F.-Y.} \bibnamefont{{Wang}}},
  \bibnamefont{and} \bibinfo{author}{\bibfnamefont{B.-B.}
  \bibnamefont{{Zhang}}}, in \emph{\bibinfo{booktitle}{Handbook of X-ray and
  Gamma-ray Astrophysics. Edited by Cosimo Bambi and Andrea Santangelo}}
  (\bibinfo{year}{2022}), p.~\bibinfo{pages}{31}.

\bibitem[{\citenamefont{{Singh} and {Desai}}(2022)}]{Singh}
\bibinfo{author}{\bibfnamefont{A.}~\bibnamefont{{Singh}}} \bibnamefont{and}
  \bibinfo{author}{\bibfnamefont{S.}~\bibnamefont{{Desai}}},
  \bibinfo{journal}{\jcap} \textbf{\bibinfo{volume}{2022}}, \bibinfo{eid}{010}
  (\bibinfo{year}{2022}), \eprint{2108.00395}.

\bibitem[{\citenamefont{{Kouveliotou} et~al.}(1993)\citenamefont{{Kouveliotou},
  {Meegan}, {Fishman}, {Bhat}, {Briggs}, {Koshut}, {Paciesas}, and
  {Pendleton}}}]{Kouveliotou}
\bibinfo{author}{\bibfnamefont{C.}~\bibnamefont{{Kouveliotou}}},
  \bibinfo{author}{\bibfnamefont{C.~A.} \bibnamefont{{Meegan}}},
  \bibinfo{author}{\bibfnamefont{G.~J.} \bibnamefont{{Fishman}}},
  \bibinfo{author}{\bibfnamefont{N.~P.} \bibnamefont{{Bhat}}},
  \bibinfo{author}{\bibfnamefont{M.~S.} \bibnamefont{{Briggs}}},
  \bibinfo{author}{\bibfnamefont{T.~M.} \bibnamefont{{Koshut}}},
  \bibinfo{author}{\bibfnamefont{W.~S.} \bibnamefont{{Paciesas}}},
  \bibnamefont{and} \bibinfo{author}{\bibfnamefont{G.~N.}
  \bibnamefont{{Pendleton}}}, \bibinfo{journal}{\apjl}
  \textbf{\bibinfo{volume}{413}}, \bibinfo{pages}{L101} (\bibinfo{year}{1993}).

\bibitem[{\citenamefont{{Woosley} and {Bloom}}(2006)}]{Woosley}
\bibinfo{author}{\bibfnamefont{S.~E.} \bibnamefont{{Woosley}}}
  \bibnamefont{and} \bibinfo{author}{\bibfnamefont{J.~S.}
  \bibnamefont{{Bloom}}}, \bibinfo{journal}{\araa}
  \textbf{\bibinfo{volume}{44}}, \bibinfo{pages}{507} (\bibinfo{year}{2006}),
  \eprint{astro-ph/0609142}.

\bibitem[{\citenamefont{{Nakar}}(2007)}]{Nakar}
\bibinfo{author}{\bibfnamefont{E.}~\bibnamefont{{Nakar}}},
  \bibinfo{journal}{\physrep} \textbf{\bibinfo{volume}{442}},
  \bibinfo{pages}{166} (\bibinfo{year}{2007}), \eprint{astro-ph/0701748}.

\bibitem[{\citenamefont{{Kulkarni} and {Desai}}(2017)}]{Kulkarni}
\bibinfo{author}{\bibfnamefont{S.}~\bibnamefont{{Kulkarni}}} \bibnamefont{and}
  \bibinfo{author}{\bibfnamefont{S.}~\bibnamefont{{Desai}}},
  \bibinfo{journal}{\apss} \textbf{\bibinfo{volume}{362}}, \bibinfo{eid}{70}
  (\bibinfo{year}{2017}), \eprint{1612.08235}.

\bibitem[{\citenamefont{{Bhave} et~al.}(2022)\citenamefont{{Bhave}, {Kulkarni},
  {Desai}, and {Srijith}}}]{Bhave}
\bibinfo{author}{\bibfnamefont{A.}~\bibnamefont{{Bhave}}},
  \bibinfo{author}{\bibfnamefont{S.}~\bibnamefont{{Kulkarni}}},
  \bibinfo{author}{\bibfnamefont{S.}~\bibnamefont{{Desai}}}, \bibnamefont{and}
  \bibinfo{author}{\bibfnamefont{P.~K.} \bibnamefont{{Srijith}}},
  \bibinfo{journal}{\apss} \textbf{\bibinfo{volume}{367}}, \bibinfo{eid}{39}
  (\bibinfo{year}{2022}), \eprint{1708.05668}.

\bibitem[{\citenamefont{{Ellis} et~al.}(2006)\citenamefont{{Ellis},
  {Mavromatos}, {Nanopoulos}, {Sakharov}, and {Sarkisyan}}}]{Ellis}
\bibinfo{author}{\bibfnamefont{J.}~\bibnamefont{{Ellis}}},
  \bibinfo{author}{\bibfnamefont{N.~E.} \bibnamefont{{Mavromatos}}},
  \bibinfo{author}{\bibfnamefont{D.~V.} \bibnamefont{{Nanopoulos}}},
  \bibinfo{author}{\bibfnamefont{A.~S.} \bibnamefont{{Sakharov}}},
  \bibnamefont{and} \bibinfo{author}{\bibfnamefont{E.~K.~G.}
  \bibnamefont{{Sarkisyan}}}, \bibinfo{journal}{Astroparticle Physics}
  \textbf{\bibinfo{volume}{25}}, \bibinfo{pages}{402} (\bibinfo{year}{2006}),
  \eprint{astro-ph/0510172}.

\bibitem[{\citenamefont{{Wei} et~al.}(2017{\natexlab{a}})\citenamefont{{Wei},
  {Zhang}, {Shao}, {Wu}, and {M{\'e}sz{\'a}ros}}}]{Wei}
\bibinfo{author}{\bibfnamefont{J.-J.} \bibnamefont{{Wei}}},
  \bibinfo{author}{\bibfnamefont{B.-B.} \bibnamefont{{Zhang}}},
  \bibinfo{author}{\bibfnamefont{L.}~\bibnamefont{{Shao}}},
  \bibinfo{author}{\bibfnamefont{X.-F.} \bibnamefont{{Wu}}}, \bibnamefont{and}
  \bibinfo{author}{\bibfnamefont{P.}~\bibnamefont{{M{\'e}sz{\'a}ros}}},
  \bibinfo{journal}{\apjl} \textbf{\bibinfo{volume}{834}}, \bibinfo{eid}{L13}
  (\bibinfo{year}{2017}{\natexlab{a}}), \eprint{1612.09425}.

\bibitem[{\citenamefont{{Du} et~al.}(2021)\citenamefont{{Du}, {Lan}, {Wei},
  {Zhou}, {Gao}, {Jiang}, {Zhang}, {Liu}, {Wu}, {Liang} et~al.}}]{Du}
\bibinfo{author}{\bibfnamefont{S.-S.} \bibnamefont{{Du}}},
  \bibinfo{author}{\bibfnamefont{L.}~\bibnamefont{{Lan}}},
  \bibinfo{author}{\bibfnamefont{J.-J.} \bibnamefont{{Wei}}},
  \bibinfo{author}{\bibfnamefont{Z.-M.} \bibnamefont{{Zhou}}},
  \bibinfo{author}{\bibfnamefont{H.}~\bibnamefont{{Gao}}},
  \bibinfo{author}{\bibfnamefont{L.-Y.} \bibnamefont{{Jiang}}},
  \bibinfo{author}{\bibfnamefont{B.-B.} \bibnamefont{{Zhang}}},
  \bibinfo{author}{\bibfnamefont{Z.-K.} \bibnamefont{{Liu}}},
  \bibinfo{author}{\bibfnamefont{X.-F.} \bibnamefont{{Wu}}},
  \bibinfo{author}{\bibfnamefont{E.-W.} \bibnamefont{{Liang}}},
  \bibnamefont{et~al.}, \bibinfo{journal}{\apj} \textbf{\bibinfo{volume}{906}},
  \bibinfo{eid}{8} (\bibinfo{year}{2021}), \eprint{2010.16029}.

\bibitem[{\citenamefont{{Ganguly} and {Desai}}(2017)}]{Ganguly}
\bibinfo{author}{\bibfnamefont{S.}~\bibnamefont{{Ganguly}}} \bibnamefont{and}
  \bibinfo{author}{\bibfnamefont{S.}~\bibnamefont{{Desai}}},
  \bibinfo{journal}{Astroparticle Physics} \textbf{\bibinfo{volume}{94}},
  \bibinfo{pages}{17} (\bibinfo{year}{2017}), \eprint{1706.01202}.

\bibitem[{\citenamefont{{Liang} et~al.}(2023)\citenamefont{{Liang}, {Lu},
  {Peng}, and {Chen}}}]{Liang23}
\bibinfo{author}{\bibfnamefont{W.-Q.} \bibnamefont{{Liang}}},
  \bibinfo{author}{\bibfnamefont{R.-J.} \bibnamefont{{Lu}}},
  \bibinfo{author}{\bibfnamefont{C.-F.} \bibnamefont{{Peng}}},
  \bibnamefont{and} \bibinfo{author}{\bibfnamefont{W.-H.}
  \bibnamefont{{Chen}}}, \bibinfo{journal}{\apj}
  \textbf{\bibinfo{volume}{942}}, \bibinfo{eid}{67} (\bibinfo{year}{2023}),
  \eprint{2212.05718}.

\bibitem[{\citenamefont{{Agrawal} et~al.}(2021)\citenamefont{{Agrawal},
  {Singirikonda}, and {Desai}}}]{Agrawal_2021}
\bibinfo{author}{\bibfnamefont{R.}~\bibnamefont{{Agrawal}}},
  \bibinfo{author}{\bibfnamefont{H.}~\bibnamefont{{Singirikonda}}},
  \bibnamefont{and} \bibinfo{author}{\bibfnamefont{S.}~\bibnamefont{{Desai}}},
  \bibinfo{journal}{\jcap} \textbf{\bibinfo{volume}{2021}}, \bibinfo{eid}{029}
  (\bibinfo{year}{2021}), \eprint{2102.11248}.

\bibitem[{\citenamefont{{Wei} and {Wu}}(2021)}]{WeiWu2021}
\bibinfo{author}{\bibfnamefont{J.-J.} \bibnamefont{{Wei}}} \bibnamefont{and}
  \bibinfo{author}{\bibfnamefont{X.-F.} \bibnamefont{{Wu}}},
  \bibinfo{journal}{Frontiers of Physics} \textbf{\bibinfo{volume}{16}},
  \bibinfo{eid}{44300} (\bibinfo{year}{2021}), \eprint{2102.03724}.

\bibitem[{\citenamefont{{Wei} and {Wu}}(2022)}]{WeiWu2}
\bibinfo{author}{\bibfnamefont{J.-J.} \bibnamefont{{Wei}}} \bibnamefont{and}
  \bibinfo{author}{\bibfnamefont{X.-F.} \bibnamefont{{Wu}}}, in
  \emph{\bibinfo{booktitle}{Handbook of X-ray and Gamma-ray Astrophysics.
  Edited by Cosimo Bambi and Andrea Santangelo}} (\bibinfo{year}{2022}),
  p.~\bibinfo{pages}{82}.

\bibitem[{\citenamefont{{Desai}}(2023)}]{Desairev}
\bibinfo{author}{\bibfnamefont{S.}~\bibnamefont{{Desai}}},
  \bibinfo{journal}{arXiv e-prints} \bibinfo{eid}{arXiv:2303.10643}
  (\bibinfo{year}{2023}), \eprint{2303.10643}.

\bibitem[{\citenamefont{{Desai} et~al.}(2023)\citenamefont{{Desai}, {Agrawal},
  and {Singirikonda}}}]{Desai23}
\bibinfo{author}{\bibfnamefont{S.}~\bibnamefont{{Desai}}},
  \bibinfo{author}{\bibfnamefont{R.}~\bibnamefont{{Agrawal}}},
  \bibnamefont{and}
  \bibinfo{author}{\bibfnamefont{H.}~\bibnamefont{{Singirikonda}}},
  \bibinfo{journal}{European Physical Journal C} \textbf{\bibinfo{volume}{83}},
  \bibinfo{eid}{63} (\bibinfo{year}{2023}), \eprint{2205.12780}.

\bibitem[{\citenamefont{{von Kienlin} et~al.}(2020)\citenamefont{{von Kienlin},
  {Meegan}, {Paciesas}, {Bhat}, {Bissaldi}, {Briggs}, {Burns}, {Cleveland},
  {Gibby}, {Giles} et~al.}}]{VonKienlin}
\bibinfo{author}{\bibfnamefont{A.}~\bibnamefont{{von Kienlin}}},
  \bibinfo{author}{\bibfnamefont{C.~A.} \bibnamefont{{Meegan}}},
  \bibinfo{author}{\bibfnamefont{W.~S.} \bibnamefont{{Paciesas}}},
  \bibinfo{author}{\bibfnamefont{P.~N.} \bibnamefont{{Bhat}}},
  \bibinfo{author}{\bibfnamefont{E.}~\bibnamefont{{Bissaldi}}},
  \bibinfo{author}{\bibfnamefont{M.~S.} \bibnamefont{{Briggs}}},
  \bibinfo{author}{\bibfnamefont{E.}~\bibnamefont{{Burns}}},
  \bibinfo{author}{\bibfnamefont{W.~H.} \bibnamefont{{Cleveland}}},
  \bibinfo{author}{\bibfnamefont{M.~H.} \bibnamefont{{Gibby}}},
  \bibinfo{author}{\bibfnamefont{M.~M.} \bibnamefont{{Giles}}},
  \bibnamefont{et~al.}, \bibinfo{journal}{\apj} \textbf{\bibinfo{volume}{893}},
  \bibinfo{eid}{46} (\bibinfo{year}{2020}), \eprint{2002.11460}.

\bibitem[{\citenamefont{{Yang} et~al.}(2020)\citenamefont{{Yang}, {Chand},
  {Zhang}, {Yang}, {Zou}, {Yang}, {Zhao}, {Shao}, {Xiong}, {Luo}
  et~al.}}]{Yang20}
\bibinfo{author}{\bibfnamefont{J.}~\bibnamefont{{Yang}}},
  \bibinfo{author}{\bibfnamefont{V.}~\bibnamefont{{Chand}}},
  \bibinfo{author}{\bibfnamefont{B.-B.} \bibnamefont{{Zhang}}},
  \bibinfo{author}{\bibfnamefont{Y.-H.} \bibnamefont{{Yang}}},
  \bibinfo{author}{\bibfnamefont{J.-H.} \bibnamefont{{Zou}}},
  \bibinfo{author}{\bibfnamefont{Y.-S.} \bibnamefont{{Yang}}},
  \bibinfo{author}{\bibfnamefont{X.-H.} \bibnamefont{{Zhao}}},
  \bibinfo{author}{\bibfnamefont{L.}~\bibnamefont{{Shao}}},
  \bibinfo{author}{\bibfnamefont{S.-L.} \bibnamefont{{Xiong}}},
  \bibinfo{author}{\bibfnamefont{Q.}~\bibnamefont{{Luo}}},
  \bibnamefont{et~al.}, \bibinfo{journal}{\apj} \textbf{\bibinfo{volume}{899}},
  \bibinfo{eid}{106} (\bibinfo{year}{2020}), \eprint{2010.05128}.

\bibitem[{\citenamefont{{Zhang} et~al.}(2021)\citenamefont{{Zhang}, {Liu},
  {Peng}, {Li}, {L{\"u}}, {Yang}, {Yang}, {Yang}, {Meng}, {Zou}
  et~al.}}]{Zhang21}
\bibinfo{author}{\bibfnamefont{B.~B.} \bibnamefont{{Zhang}}},
  \bibinfo{author}{\bibfnamefont{Z.~K.} \bibnamefont{{Liu}}},
  \bibinfo{author}{\bibfnamefont{Z.~K.} \bibnamefont{{Peng}}},
  \bibinfo{author}{\bibfnamefont{Y.}~\bibnamefont{{Li}}},
  \bibinfo{author}{\bibfnamefont{H.~J.} \bibnamefont{{L{\"u}}}},
  \bibinfo{author}{\bibfnamefont{J.}~\bibnamefont{{Yang}}},
  \bibinfo{author}{\bibfnamefont{Y.~S.} \bibnamefont{{Yang}}},
  \bibinfo{author}{\bibfnamefont{Y.~H.} \bibnamefont{{Yang}}},
  \bibinfo{author}{\bibfnamefont{Y.~Z.} \bibnamefont{{Meng}}},
  \bibinfo{author}{\bibfnamefont{J.~H.} \bibnamefont{{Zou}}},
  \bibnamefont{et~al.}, \bibinfo{journal}{Nature Astronomy}
  \textbf{\bibinfo{volume}{5}}, \bibinfo{pages}{911} (\bibinfo{year}{2021}),
  \eprint{2105.05021}.

\bibitem[{\citenamefont{{Wang} et~al.}(2021)\citenamefont{{Wang}, {Zheng},
  {Xiao}, {Yang}, {Liu}, {Yang}, {Zou}, {Zhang}, {Zeng}, {Xiong}
  et~al.}}]{Wang21}
\bibinfo{author}{\bibfnamefont{X.~I.} \bibnamefont{{Wang}}},
  \bibinfo{author}{\bibfnamefont{X.}~\bibnamefont{{Zheng}}},
  \bibinfo{author}{\bibfnamefont{S.}~\bibnamefont{{Xiao}}},
  \bibinfo{author}{\bibfnamefont{J.}~\bibnamefont{{Yang}}},
  \bibinfo{author}{\bibfnamefont{Z.-K.} \bibnamefont{{Liu}}},
  \bibinfo{author}{\bibfnamefont{Y.-H.} \bibnamefont{{Yang}}},
  \bibinfo{author}{\bibfnamefont{J.-H.} \bibnamefont{{Zou}}},
  \bibinfo{author}{\bibfnamefont{B.-B.} \bibnamefont{{Zhang}}},
  \bibinfo{author}{\bibfnamefont{M.}~\bibnamefont{{Zeng}}},
  \bibinfo{author}{\bibfnamefont{S.-L.} \bibnamefont{{Xiong}}},
  \bibnamefont{et~al.}, \bibinfo{journal}{\apj} \textbf{\bibinfo{volume}{922}},
  \bibinfo{eid}{237} (\bibinfo{year}{2021}), \eprint{2107.10452}.

\bibitem[{\citenamefont{{Wei} et~al.}(2017{\natexlab{b}})\citenamefont{{Wei},
  {Wu}, {Zhang}, {Shao}, {M{\'e}sz{\'a}ros}, and {Kosteleck{\'y}}}}]{Wei2}
\bibinfo{author}{\bibfnamefont{J.-J.} \bibnamefont{{Wei}}},
  \bibinfo{author}{\bibfnamefont{X.-F.} \bibnamefont{{Wu}}},
  \bibinfo{author}{\bibfnamefont{B.-B.} \bibnamefont{{Zhang}}},
  \bibinfo{author}{\bibfnamefont{L.}~\bibnamefont{{Shao}}},
  \bibinfo{author}{\bibfnamefont{P.}~\bibnamefont{{M{\'e}sz{\'a}ros}}},
  \bibnamefont{and} \bibinfo{author}{\bibfnamefont{V.~A.}
  \bibnamefont{{Kosteleck{\'y}}}}, \bibinfo{journal}{\apj}
  \textbf{\bibinfo{volume}{842}}, \bibinfo{eid}{115}
  (\bibinfo{year}{2017}{\natexlab{b}}), \eprint{1704.05984}.

\bibitem[{\citenamefont{{Wei} and {Wu}}(2017)}]{Wei17}
\bibinfo{author}{\bibfnamefont{J.-J.} \bibnamefont{{Wei}}} \bibnamefont{and}
  \bibinfo{author}{\bibfnamefont{X.-F.} \bibnamefont{{Wu}}},
  \bibinfo{journal}{\apj} \textbf{\bibinfo{volume}{851}}, \bibinfo{eid}{127}
  (\bibinfo{year}{2017}), \eprint{1711.09185}.

\bibitem[{\citenamefont{{Pan} et~al.}(2020)\citenamefont{{Pan}, {Qi}, {Cao},
  {Liu}, {Liu}, {Geng}, {Lian}, and {Zhu}}}]{Pan}
\bibinfo{author}{\bibfnamefont{Y.}~\bibnamefont{{Pan}}},
  \bibinfo{author}{\bibfnamefont{J.}~\bibnamefont{{Qi}}},
  \bibinfo{author}{\bibfnamefont{S.}~\bibnamefont{{Cao}}},
  \bibinfo{author}{\bibfnamefont{T.}~\bibnamefont{{Liu}}},
  \bibinfo{author}{\bibfnamefont{Y.}~\bibnamefont{{Liu}}},
  \bibinfo{author}{\bibfnamefont{S.}~\bibnamefont{{Geng}}},
  \bibinfo{author}{\bibfnamefont{Y.}~\bibnamefont{{Lian}}}, \bibnamefont{and}
  \bibinfo{author}{\bibfnamefont{Z.-H.} \bibnamefont{{Zhu}}},
  \bibinfo{journal}{\apj} \textbf{\bibinfo{volume}{890}}, \bibinfo{eid}{169}
  (\bibinfo{year}{2020}), \eprint{2001.08451}.

\bibitem[{\citenamefont{{Shao} et~al.}(2017)\citenamefont{{Shao}, {Zhang},
  {Wang}, {Wu}, {Cheng}, {Zhang}, {Yu}, {Xi}, {Wang}, {Feng} et~al.}}]{Shao}
\bibinfo{author}{\bibfnamefont{L.}~\bibnamefont{{Shao}}},
  \bibinfo{author}{\bibfnamefont{B.-B.} \bibnamefont{{Zhang}}},
  \bibinfo{author}{\bibfnamefont{F.-R.} \bibnamefont{{Wang}}},
  \bibinfo{author}{\bibfnamefont{X.-F.} \bibnamefont{{Wu}}},
  \bibinfo{author}{\bibfnamefont{Y.-H.} \bibnamefont{{Cheng}}},
  \bibinfo{author}{\bibfnamefont{X.}~\bibnamefont{{Zhang}}},
  \bibinfo{author}{\bibfnamefont{B.-Y.} \bibnamefont{{Yu}}},
  \bibinfo{author}{\bibfnamefont{B.-J.} \bibnamefont{{Xi}}},
  \bibinfo{author}{\bibfnamefont{X.}~\bibnamefont{{Wang}}},
  \bibinfo{author}{\bibfnamefont{H.-X.} \bibnamefont{{Feng}}},
  \bibnamefont{et~al.}, \bibinfo{journal}{\apj} \textbf{\bibinfo{volume}{844}},
  \bibinfo{eid}{126} (\bibinfo{year}{2017}), \eprint{1610.07191}.

\bibitem[{\citenamefont{{Jacob} and {Piran}}(2008)}]{Jacob}
\bibinfo{author}{\bibfnamefont{U.}~\bibnamefont{{Jacob}}} \bibnamefont{and}
  \bibinfo{author}{\bibfnamefont{T.}~\bibnamefont{{Piran}}},
  \bibinfo{journal}{\jcap} \textbf{\bibinfo{volume}{1}}, \bibinfo{eid}{031}
  (\bibinfo{year}{2008}), \eprint{0712.2170}.

\bibitem[{\citenamefont{{Biesiada} and {Pi{\'o}rkowska}}(2009)}]{Biesiada}
\bibinfo{author}{\bibfnamefont{M.}~\bibnamefont{{Biesiada}}} \bibnamefont{and}
  \bibinfo{author}{\bibfnamefont{A.}~\bibnamefont{{Pi{\'o}rkowska}}},
  \bibinfo{journal}{Classical and Quantum Gravity}
  \textbf{\bibinfo{volume}{26}}, \bibinfo{eid}{125007} (\bibinfo{year}{2009}),
  \eprint{1008.2615}.

\bibitem[{\citenamefont{{Planck Collaboration}
  et~al.}(2020)\citenamefont{{Planck Collaboration}, {Aghanim}, {Akrami},
  {Ashdown}, {Aumont}, {Baccigalupi}, {Ballardini}, {Banday}, {Barreiro},
  {Bartolo} et~al.}}]{Planck20}
\bibinfo{author}{\bibnamefont{{Planck Collaboration}}},
  \bibinfo{author}{\bibfnamefont{N.}~\bibnamefont{{Aghanim}}},
  \bibinfo{author}{\bibfnamefont{Y.}~\bibnamefont{{Akrami}}},
  \bibinfo{author}{\bibfnamefont{M.}~\bibnamefont{{Ashdown}}},
  \bibinfo{author}{\bibfnamefont{J.}~\bibnamefont{{Aumont}}},
  \bibinfo{author}{\bibfnamefont{C.}~\bibnamefont{{Baccigalupi}}},
  \bibinfo{author}{\bibfnamefont{M.}~\bibnamefont{{Ballardini}}},
  \bibinfo{author}{\bibfnamefont{A.~J.} \bibnamefont{{Banday}}},
  \bibinfo{author}{\bibfnamefont{R.~B.} \bibnamefont{{Barreiro}}},
  \bibinfo{author}{\bibfnamefont{N.}~\bibnamefont{{Bartolo}}},
  \bibnamefont{et~al.}, \bibinfo{journal}{\aap} \textbf{\bibinfo{volume}{641}},
  \bibinfo{eid}{A6} (\bibinfo{year}{2020}), \eprint{1807.06209}.

\bibitem[{\citenamefont{{Trotta}}(2017)}]{Trotta}
\bibinfo{author}{\bibfnamefont{R.}~\bibnamefont{{Trotta}}},
  \bibinfo{journal}{ArXiv e-prints}  (\bibinfo{year}{2017}),
  \eprint{1701.01467}.

\bibitem[{\citenamefont{{Kerscher} and {Weller}}(2019)}]{Weller}
\bibinfo{author}{\bibfnamefont{M.}~\bibnamefont{{Kerscher}}} \bibnamefont{and}
  \bibinfo{author}{\bibfnamefont{J.}~\bibnamefont{{Weller}}},
  \bibinfo{journal}{SciPost Physics Lecture Notes} \textbf{\bibinfo{volume}{9}}
  (\bibinfo{year}{2019}), \eprint{1901.07726}.

\bibitem[{\citenamefont{{Sharma}}(2017)}]{Sanjib}
\bibinfo{author}{\bibfnamefont{S.}~\bibnamefont{{Sharma}}},
  \bibinfo{journal}{\araa} \textbf{\bibinfo{volume}{55}}, \bibinfo{pages}{213}
  (\bibinfo{year}{2017}), \eprint{1706.01629}.

\bibitem[{\citenamefont{{Krishak} and {Desai}}(2020)}]{Krishak4}
\bibinfo{author}{\bibfnamefont{A.}~\bibnamefont{{Krishak}}} \bibnamefont{and}
  \bibinfo{author}{\bibfnamefont{S.}~\bibnamefont{{Desai}}},
  \bibinfo{journal}{\jcap} \textbf{\bibinfo{volume}{2020}}, \bibinfo{eid}{006}
  (\bibinfo{year}{2020}), \eprint{2003.10127}.

\bibitem[{\citenamefont{{Speagle}}(2020)}]{dynesty}
\bibinfo{author}{\bibfnamefont{J.~S.} \bibnamefont{{Speagle}}},
  \bibinfo{journal}{\mnras}  (\bibinfo{year}{2020}), \eprint{1904.02180}.

\bibitem[{\citenamefont{{Gunapati} et~al.}(2022)\citenamefont{{Gunapati},
  {Jain}, {Srijith}, and {Desai}}}]{Gunapati}
\bibinfo{author}{\bibfnamefont{G.}~\bibnamefont{{Gunapati}}},
  \bibinfo{author}{\bibfnamefont{A.}~\bibnamefont{{Jain}}},
  \bibinfo{author}{\bibfnamefont{P.~K.} \bibnamefont{{Srijith}}},
  \bibnamefont{and} \bibinfo{author}{\bibfnamefont{S.}~\bibnamefont{{Desai}}},
  \bibinfo{journal}{\pasa} \textbf{\bibinfo{volume}{39}}, \bibinfo{eid}{e001}
  (\bibinfo{year}{2022}), \eprint{1803.06473}.

\bibitem[{\citenamefont{{Will}}(2006)}]{Will05}
\bibinfo{author}{\bibfnamefont{C.~M.} \bibnamefont{{Will}}},
  \bibinfo{journal}{Annalen der Physik} \textbf{\bibinfo{volume}{15}},
  \bibinfo{pages}{19} (\bibinfo{year}{2006}), \eprint{gr-qc/0504086}.

\end{thebibliography}

\end{document}